# Deciphering Delivery Rider Mobility: A City-Scale, Path-Reconstructed On-Demand Delivery Route Dataset

**Chengbo Zhang** [a]

chengbo.zhang@mail.mcgill.ca

**Zuopeng Xiao** [b*] **(Corresponding Author)**

tacxzp@foxmail.com

Guangdong Provincial Humanities and Social Sciences Research Center for Digital Space Governance Technology, Guangdong, China, Shenzhen, 518033, Guangdong, China

[a] **Department of Civil Engineering, McGill University,** Montreal, Canada

[b] **Institute of Future Human Habitats, Shenzhen International Graduate School, Tsinghua University**, Shenzhen, China

## Acknowledgements

This research is funded by the National Natural Science Foundation of China (No. 42571286) and the Guangdong Provincial University Young Innovative Talents Project (2023WQNCX140).

## Abstract

The rapid expansion of the on-demand economy has profoundly reshaped urban mobility and logistics, yet open data linking multi-stop delivery task sequences with route geometry remain scarce. Here, we present a city-scale, path-reconstructed route dataset for instant-delivery tasks in Beijing, built from desensitized platform records from Ele.me. The dataset contains 79,648 reconstructed delivery-wave records associated with 986 anonymized courier identifiers and covering 267,529 orders during February 2020. For each wave, Assign, Pickup, and Delivery action points are ordered by source-record timestamps, and Amap cycling routes are queried between consecutive action points to create continuous route geometries.

The resulting dataset records navigation-based riding paths and the associated action sequence for each delivery wave, revealing insightful spatiotemporal regularities. Source-record validation shows strong agreement for distance and moderate agreement for duration. This urban data resource enables researchers in urban analytics and transportation management to investigate delivery activities, model urban logistics systems, and develop sustainable policies.

## Introduction

The rapid expansion of instant e-commerce has profoundly reshaped urban logistics, introducing a significant new mobility component: the instant delivery rider (Krier et al., 2022). This global workforce, estimated at over 25 million individuals (Backlinko Team, 2025; Dong, 2024; Kumar, 2025), powers a market exceeding $380 billion in 2024 (Grand View Research, 2023). The sector's explosive growth underscores its surging impact while raising profound sustainability challenges (Lord et al., 2023). A granular understanding of rider mobility is therefore essential for optimizing logistics, developing sustainable transport policies, and modeling interactions between delivery traffic and general urban congestion.

However, empirical research on instant-delivery routing remains constrained by limited access to open route-level representations of multi-stop delivery tasks. Due to commercial sensitivity and privacy concerns, datasets are typically proprietary, forcing researchers to rely on questionnaires or small-scale field observations (Oviedo-Trespalacios et al., 2022; Vecchio et al., 2022; Yu et al., 2024). While the increasing availability of general human mobility datasets is widening urban research avenues (Yabe et al., 2024), a critical scarcity of data specific to delivery riders persists. This data gap hinders the scientific community's ability to monitor, model, and develop sustainable management strategies for this transformative industry. Although some datasets containing origin-destination (OD) information for delivery orders are available (Qiu et al., 2021), accurately reconstructing the continuous, multi-stop trajectories of riders from such sparse data remains a significant challenge.

Accurately modeling instant delivery mobility necessitates incorporating the order wave—a complete, multi-stop tour assigned to a single rider—rather than relying on disconnected OD pairs. In practice, a wave (also called a bundle) is defined by a specific sequence of designated tasks (e.g., Action Point 1: pick up *Order A*; Action Point 2: pick up *Order B*; Action Point 3: deliver *Order A*; Action Point 4: deliver *Order B*), nonlinearly linking multiple orders within a continuous work period to manage supply-demand surges (Zhu et al., 2020). While agent-based models attempt to capture these travel dynamics (Kashiyama et al., 2022), they often struggle with calibration and operational realism during on-demand delivery. In contrast, online map services serve as indispensable tools for actual platform-based work, offering a lightweight approach that reflects real-world road conditions (Toledo et al., 2020; Xiao et al., 2025). Consequently, our framework leverages this by sequentially reconstructing cycling-optimized routes between each consecutive action point, generating a single, continuous trajectory aligned with realistic physical and operational constraints.

In this paper, we present a city-scale, path-reconstructed route dataset for on-demand instant delivery riders in Beijing, China. The dataset is derived from a 28-day public Ele.me/Tianchi release of desensitized delivery-task records, including order, courier, distance, and action tables. We first reconstruct wave-based delivery chains from the source records and then query Amap cycling routes between consecutive Action Points in each wave. The resulting dataset contains 79,648 reconstructed delivery-wave records associated with 986 anonymized courier identifiers and covering 267,529 orders. Its main value lies in representing the complex, multi-stop task-sequence structure of delivery waves at street-network resolution: 60.62% of waves include multiple pickup actions before the first delivery, and 78.05% of multi-order waves do not follow a simple pickup-delivery alternating sequence.

This dataset supports the following research directions and applications, subject to its navigation-based path-reconstruction assumptions:

- Spatiotemporal delivery-routing patterns: The dataset supports city-scale analysis of delivery corridors, pickup and delivery hotspots, wave sequencing, and route-network coverage, providing evidence for transport-geography research on platform-mediated delivery activity.
- Operational capacity and workload: Action sequences, wave sizes and rider activity allow researchers to quantify bundled task structures, courier action, temporal delivery capacity variation, and categorize delivery styles.
- Transport-environment analysis: Delivery routes can be linked with road-network, land-use, emissions-factor, traffic-exposure, or safety-exposure data for network-level transport-environment analysis.

## Delivery Route Reconstruction Framework

### *Delivery record datasets*

The foundational data used in this study were obtained from the Smart Logistics competition, a public data science challenge hosted on the Alibaba Tianchi platform in 2020 (https://tianchi.aliyun.com/competition/entrance/231777/introduction). According to the official data documentation, the dataset contains anonymized records of historical delivery tasks in Beijing, organized into multiple data sheets, including wave-level information on multi-order tours assigned to individual riders. Personal identifiers such as real rider names or contact information were removed before release. Our reconstruction uses anonymized task identifiers, coordinates, behavior timestamps, and action categories from these source tables. Therefore, the published dataset contains no personal rider identities or continuous GPS traces outside of the active delivery process.

### *Path-reconstruction framework*

Our path-reconstruction framework is designed to generate a continuous, high-resolution route for each multi-stop delivery wave (**Fig.** 1). The core principle is to simulate the most probable path a rider would take to complete the sequence of tasks within a given delivery tour, transforming discrete task locations into a reconstructed route representation that is consistent with the recorded task sequence and street-network constraints.

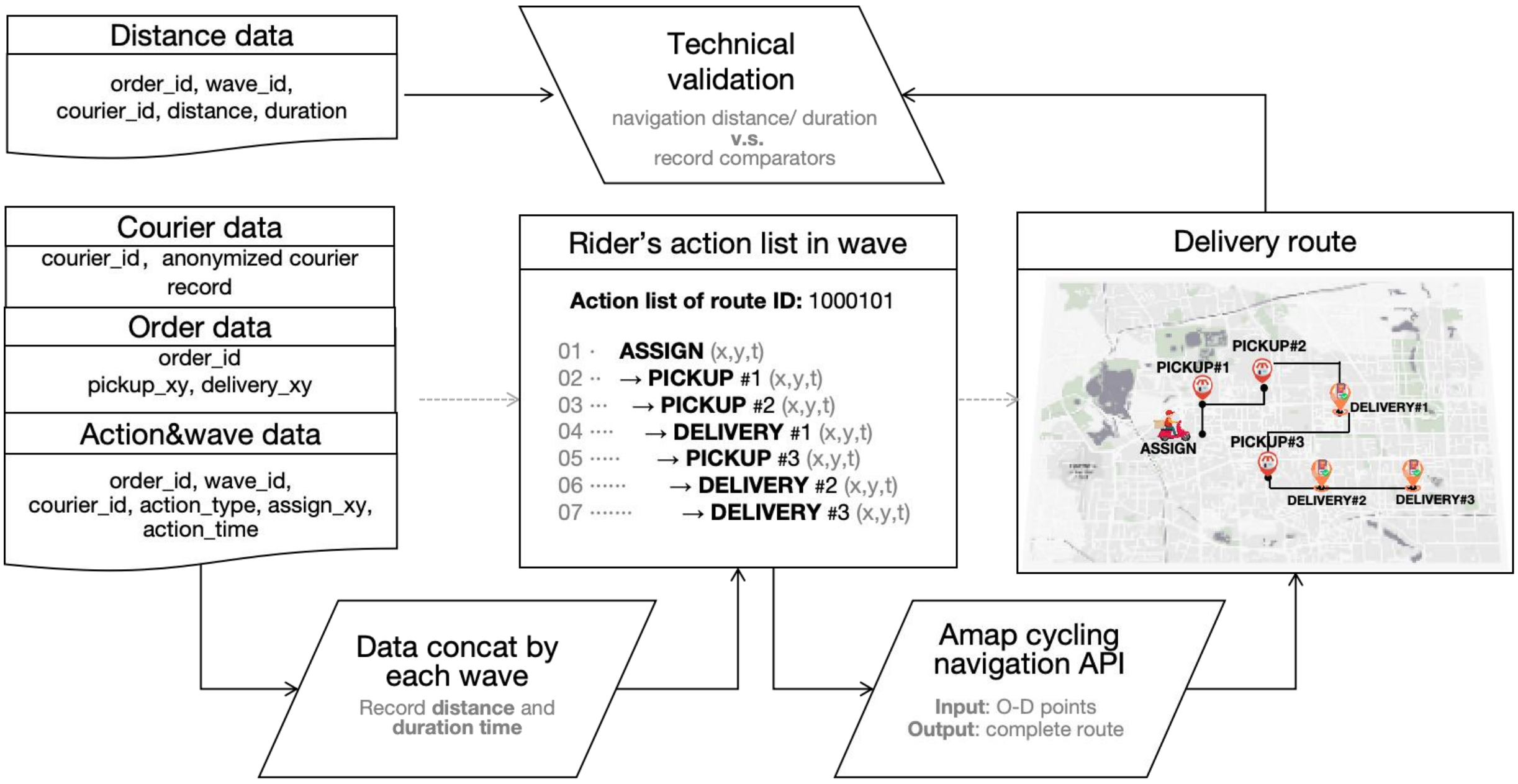

**Fig. 1** | Workflow of data processing and delivery route generation.

The process begins with the decomposition of each delivery wave from the source data. A wave, uniquely identified by its *courier_id* and *wave_index*, represents the fundamental unit of a rider's tour. Typically, a delivery wave commences at the moment of assignment and concludes upon the final delivery of the last order within the bundled tasks.

First, for each wave, the "assign," "pickup," and "delivery" tasks are sorted chronologically based on their recorded expect_time timestamps, which denote the time of the corresponding recorded behavior in the official data description. This establishes the recorded Action Point sequence used for route reconstruction. From this ordered sequence, the framework defines a series of point-to-point routing tasks, where the destination of task $i$ becomes the origin for the journey to task $i+1$. To enable subsequent evaluation of our generated routes, we also recorded the total duration and grid distance for each wave, as provided in the source data, to serve as comparable metrics.

With the sequence of Action Point pairs established for a wave, the framework then systematically queries the Amap routing Application Programming Interface (API) to reconstruct a navigation-based cycling path between consecutive Action Points under street-network constraints. Each OD pair is submitted to the API with the query explicitly configured to generate a route optimized for cycling, producing route geometry for the corresponding task-to-task segment.

Finally, the output route from the navigation service is synthesized into a standardized line-geometry format. For each path segment, the service returns a high-resolution polyline, and timestamps are interpolated along this path based on the travel time estimated by the API. This routing and synthesis process is iteratively applied to all consecutive tasks in the wave. The resulting path segments are then concatenated in order, forming a single, complete, and continuous route that represents the rider's entire multi-stop delivery tour.

## Dataset Description and Usage Note

The route dataset is provided in GeoJSON format, a standard open format for encoding geographic data structures, and is available on Figshare (https://doi.org/10.6084/m9.figshare.29314796.v1). The file contains a FeatureCollection composed of two distinct feature types, identified by the feature_type property:

- Route Features (feature_type: 'route'): Each line feature represents a complete, multi-stop delivery tour for a single rider. The properties for route features are detailed in **Table** S1.
- Point Features (feature_type: 'action_point'): Each point feature represents a specific task (e.g., ASSIGN, PICKUP, DELIVERY) within a delivery tour. The properties for point features are detailed in Table S2.

Detailed descriptions of the data records, including schemas for Route Features and Point Features, are provided in the Supplementary Materials (**Table** S1 and **Table** S2). To assess the technical quality of the dataset, we compare reconstructed route metrics with source-record comparator metrics. The reconstructed distance shows strong correspondence with the source-record distance (Pearson's r = 0.92; median absolute percent error = 16.85%; 90th-percentile (p90) = 51.58%), while duration shows moderate correspondence (Pearson's r = 0.79; median absolute percent error = 25.32%; p90 = 60.13%) (Fig. S1). Distance errors decline with route length, with median absolute percent error decreasing from 47.19% for routes below 1 km to 12.94% for routes above 10 km. Duration should therefore be interpreted as a rough travel-time comparator because source-record wave duration may include waiting, preparation, and operational delay, whereas Amap nav_dur estimates cycling travel time. Internal temporal patterns further show plausible urban delivery rhythms, including bimodal activity peaks during dining hours and weekday-weekend variation in courier capacity and workload (**Fig.** S2-S5).

To illustrate the spatiotemporal distribution of the dataset, we visualize routes from February 15th, a typical mid-month weekend day with near-median levels of active couriers, orders, and Action Points, in Fig. 2. The main panel reveals dense route coverage within Beijing's central area, showing strong alignment with the urban street network. To demonstrate the multi-stop structure of delivery waves, four individual waves are highlighted; each example connects the ordered sequence of ASSIGN, PICKUP, and DELIVERY Action Points, with timelines showing the timing and duration of each step. Kernel density plots of pickup and delivery locations further identify clear hotspots in the urban core, reflecting the spatial concentration of instant-delivery activity. A companion 3D interactive visualization of the same path-reconstructed delivery routes is also available for exploratory inspection (Zhang, 2026).

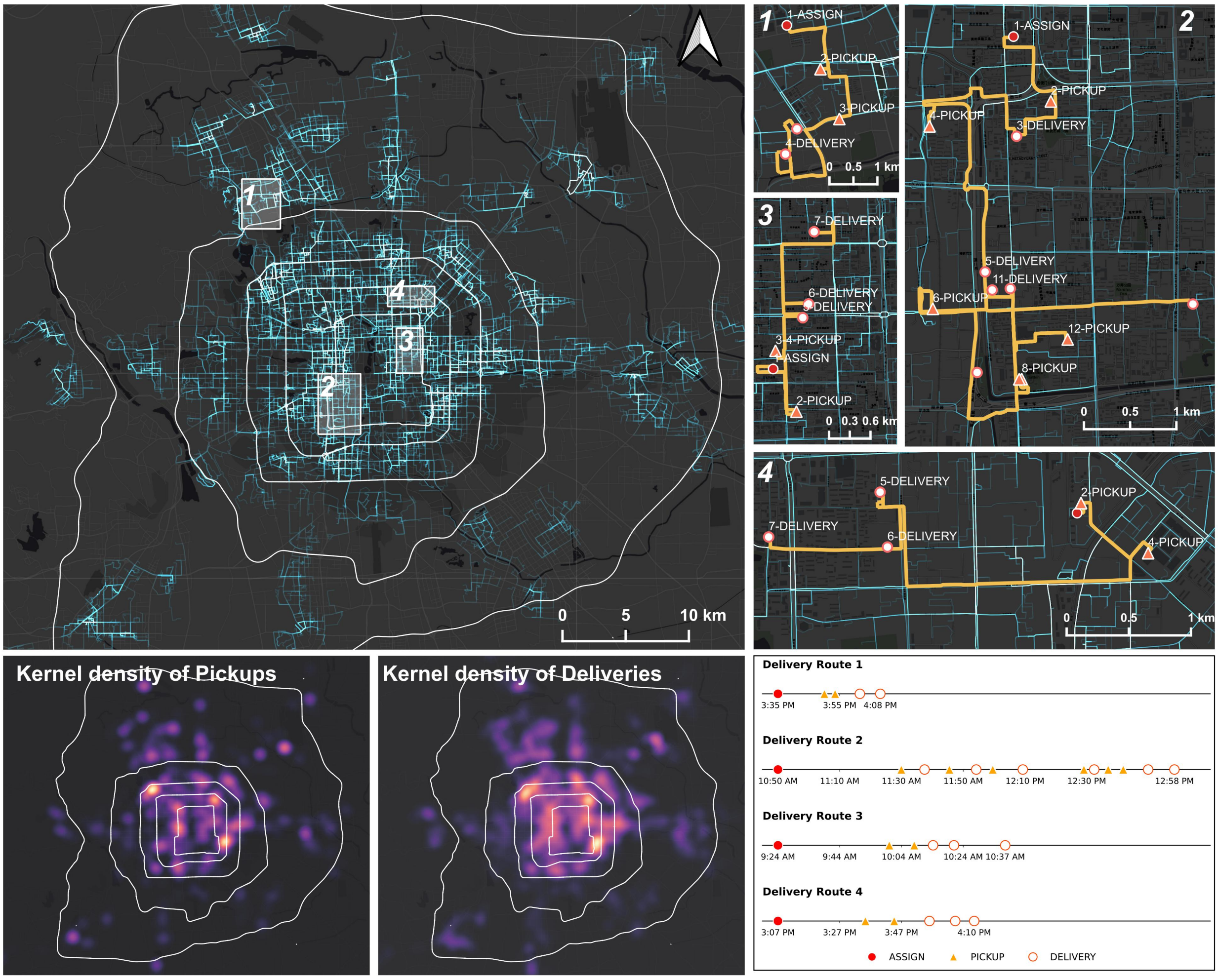

**Fig. 2** | Spatiotemporal patterns of on-demand delivery in Beijing. The main panel displays reconstructed routes from February 15, an illustrative mid-month weekend day with near-median riders/orders/action-point counts, showing dense coverage aligned with the urban street network. Insets (1 − 4) detail individual multi-stop waves and their action sequences, accompanied by timelines illustrating step durations. Kernel density plots (bottom left) reveal spatial hotspots for pickup and delivery activities.

The data are publicly available under the CC BY 4.0 license. For programmatic analysis, we recommend using Python with geopandas to load the data as GeoDataFrames, enabling geometric operations via shapely while preserving metadata. Several technical aspects should be considered when using the dataset. First, the source records cover February 1 − 28, 2020, during the early COVID-19 period in Beijing, when restaurant operations, household activity patterns, rider supply, and delivery demand may have differed from those in non-pandemic periods. Second, each route is limited to a single delivery wave and does not include travel before assignment or after the final delivery. Third, the dataset is designed for analyzing reconstructed delivery-route geometry, Action Point sequencing, and city-scale delivery-network patterns. Fine-scale route-choice behavior, detours, sidewalk riding, and safety exposure still require direct empirical observations as complements.

## Conclusions

This article presents a city-scale, path-reconstructed route dataset that elucidates the granular mobility patterns of on-demand instant delivery riders in Beijing. By linking desensitized platform task records with cycling routes generated from a major online navigation service, the dataset converts sparse action point sequences into continuous route geometries for 79,648 reconstructed delivery-wave records. The resulting resource helps bridge the gap between order-level records and network-based representations of bundled delivery work. Metric validation indicates strong agreement for distance and moderate agreement for duration relative to the source-record metrics, supporting city-scale spatial routing analysis while motivating cautious interpretation of time-sensitive applications and short-distance precision. Future work can extend this framework by incorporating observed GPS traces, dwell-time models, dispatch-state records, and post-pandemic datasets to evaluate route choice, waiting time, and operational adaptation more directly.

## Code availability

The Python code for processing original data and generating routes by Amap API is available on GitHub (https://github.com/Nicholas0027/ODIDMobTraj).

# Supplementary Material

## Data Records

**Table S1**. Schema of Route Features

| Field Name | Data Type | Description |
|---|---|---|
| Route_id | Integer | A unique, consecutive identifier for each delivery route across the entire dataset. |
| courier_id | Integer | Anonymized identifier for the delivery rider. |
| date | String (YYYY-MM-DD) | The date on which the delivery route occurred. |
| no_act | Integer | The total number of actions (stops) within the route. |
| act_lst | String (List) | A string representation of the chronological list of action types (e.g., "['ASSIGN', 'PICKUP', 'DELIVERY']"). |
| r_time_lst | String (List) | A list of source-record UNIX timestamps for ASSIGN, PICKUP, and DELIVERY Action Points within the route. |
| r_dis_lst | String (List) | A list of source-record segment distance metrics inherited from the distance table. |
| r_dur_all | Integer | The source-record wave duration in seconds, computed from recorded behavior timestamps as the time from assignment to the last delivery action. |
| r_dis_all | Float | The sum of source-record segment distance metrics in meters. |
| no_nav | Integer | The number of navigation segments that constitute the route. |
| nav_dis | Float | Total Amap cycling-route distance for the reconstructed route, in meters. |
| nav_dur | Integer | Total Amap cycling-route travel time for the reconstructed route, in seconds. |
| rider_lvl | Integer | The experience level of the rider. |
| rider_spd | Float | The rider's recorded average speed. |
| max_load | Integer | The maximum load capacity of the rider. |
| wthr_grd | Integer | The grade of the weather conditions during the route. |
| feature_type | String | A constant value: 'route'. |

**Table S2.** Schema of Action Point Features

| Field Name | Data Type | Description |
|---|---|---|
| Route_id | Integer | The identifier of the route to which this point belongs. Links points to their parent route. |
| act_pt_id | String | A unique 7-digit identifier for each action point, formed by combining the 5-digit Route_id and the 2-digit act_order. |
| courier_id | Integer | Anonymized identifier for the delivery rider. |
| date | String (YYYY-MM-DD) | The date on which the action occurred. |
| act_time | Integer | The UNIX timestamp for when the action occurred. |
| act_order | Integer | The chronological sequence number (starting from 1) of the action within its route. |
| action_type | String | The type of action (e.g., 'ASSIGN', 'PICKUP', or 'DELIVERY'). |
| feature_type | String | A constant value: 'action_point'. |

## Technical Validation

We assessed the internal consistency of the path-reconstructed delivery route dataset using two complementary checks. First, we compared the reconstructed route metrics with source-record comparator metrics available for each delivery wave. These comparator metrics were inherited from or derived from the platform records and were used in lieu of unavailable ground-truth GPS trajectories. Second, we examined the internal temporal patterns of action points and courier identifiers to evaluate whether the reconstructed dataset preserves interpretable delivery-demand rhythms.

### Validation Against Source-Record Metrics

The primary comparator check evaluates two pairs of variables: (1) source-record route distance (r_dis_all) and reconstructed route distance (nav_dis), and (2) source-record wave duration (r_dur_all) and reconstructed route travel time (nav_dur). Together, these comparator metrics from the source-record fields provide an internal consistency check between the reconstructed routes and the platform records in the absence of ground-truth GPS trajectories.

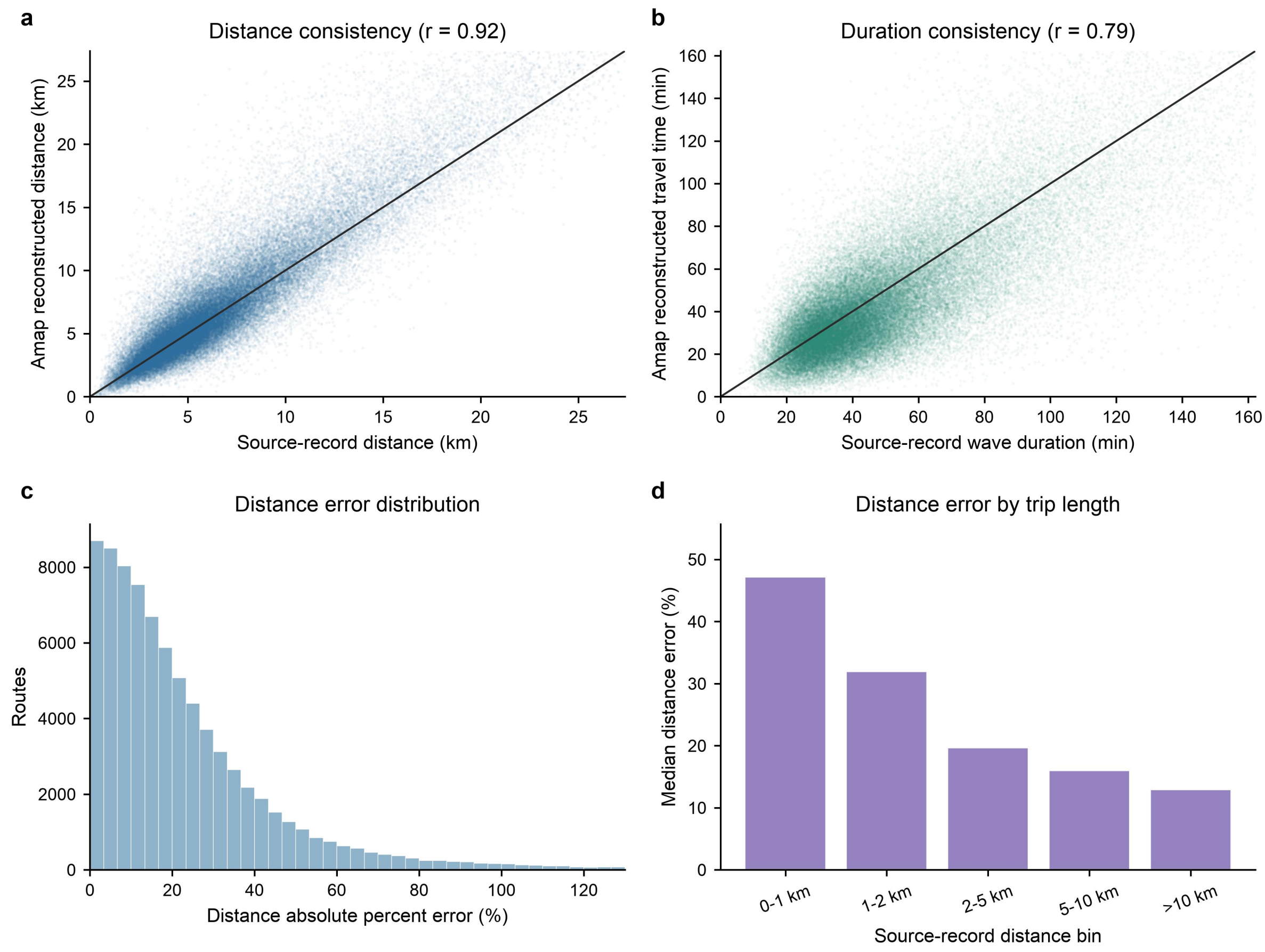

**Fig. S1** | Source-record validation of reconstructed delivery route metrics. a. Comparison between source-record distance and navigation-reconstructed route distance. b. Comparison between source-record wave duration and navigation-reconstructed cycling travel time. c. Distribution of the absolute percentage error in reconstructed route distance. d. Median distance error by source-record trip-length bin.

The validation indicates stronger agreement for distance than for duration. For distance, Pearson's r is 0.92 and Spearman's rho is 0.88. The median absolute percent error is 16.85%, and the p90 absolute percent error is 51.58%. The error pattern also varies systematically with route length. Short routes exhibit larger relative errors, with a median absolute percentage error of 47.19% for routes shorter than 1 km. The median error decreases to 31.94% for routes of 1–2 km, 19.69% for routes of 2–5 km, 16.03% for routes of 5–10 km, and 12.94% for routes longer than 10 km. This pattern is expected because small absolute differences in reconstructed distance can translate into large percentage errors for very short routes.

Duration shows weaker agreement. Pearson's r is 0.79 and Spearman's rho is 0.71, with a median absolute percent error of 25.32% and a p90 absolute percent error of 60.13%. These results suggest that duration should be used as a rough travel-time comparison rather than as a direct measure of routed cycling time. The source-record wave duration is calculated from elapsed action time within the delivery wave, whereas map navigation based nav_dur estimates cycling travel time along the reconstructed route. The source-record measure can therefore include waiting at pickup locations, restaurant preparation time, handover delays, and other operational pauses that are not represented in the routing estimate.

### *Spatiotemporal patterns of delivery action chains*

Furthermore, we validated the dataset by examining its internal temporal patterns. First, we analyzed the daily distribution of order pickups across the 28-day period, distinguishing between weekdays and weekends (**Fig**. S2). The data reveal both daily fluctuations and a general upward trend in delivery activity throughout the month. Second, we analyzed the average hourly distribution of the three main action types, separating weekdays from weekends (**Fig.** S3). The results for both periods reveal a distinct bimodal distribution, with pronounced peaks corresponding to the characteristic lunch (11:00 − 13:00) and dinner (17:00 − 19:00) rushes of on-demand food delivery. Within these rushes, the temporal ordering follows a logical sequence: ASSIGN actions peak with PICKUP actions, followed by a slightly delayed DELIVERY peak. Notably, weekend activity exhibits more moderate fluctuations than weekday activity, which shows sharper and more concentrated peaks.

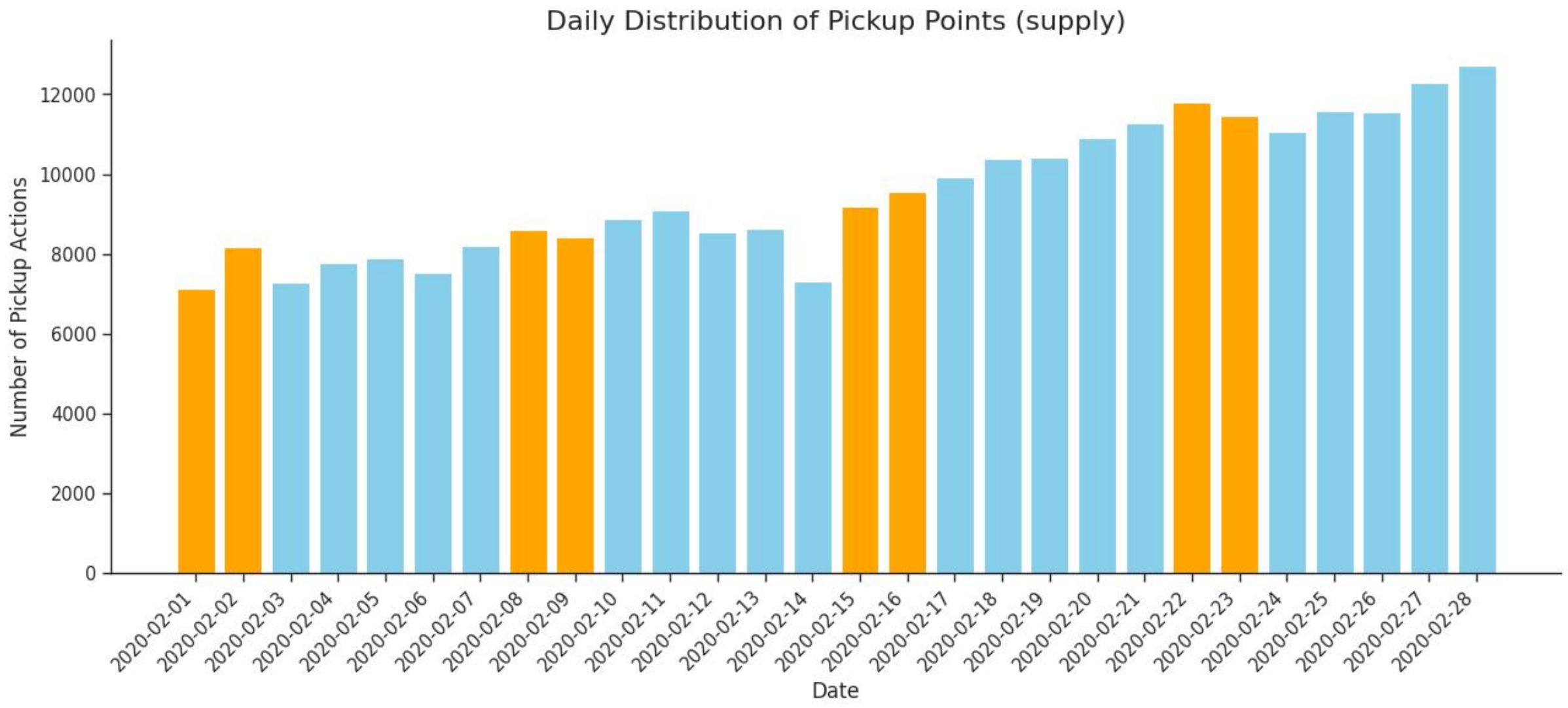

**Fig. S2** | Daily distribution of pickup actions with weekend days highlighted in orange. The plot illustrates daily fluctuations and a general increase in the number of delivery orders over the month in the dataset.

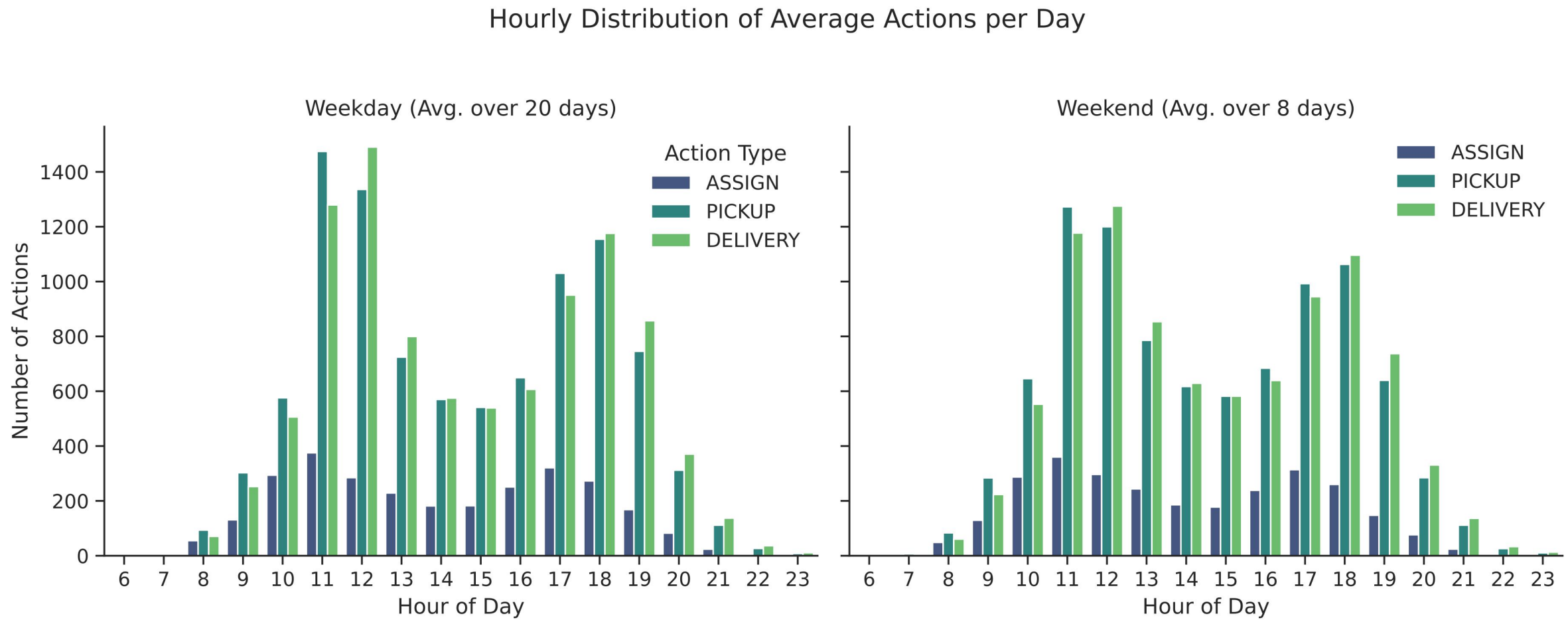

**Fig. S3** | Average hourly distribution of delivery actions on weekdays versus weekends. The plots show the average number of ASSIGN, PICKUP, and DELIVERY actions per hour, aggregated over 20 weekdays (left) and 8 weekend days (right).

In addition, we analyzed the hourly distribution of rider capacity and workload, separating weekdays from weekends to identify distinct temporal patterns. **Fig.** S4 shows the average number of active riders throughout the day. On both weekdays and weekends, rider availability forms clear peaks in labor participation during lunch and dinner hours. The comparison between the panels reveals differences in workforce availability and scheduling patterns between weekdays and weekends.

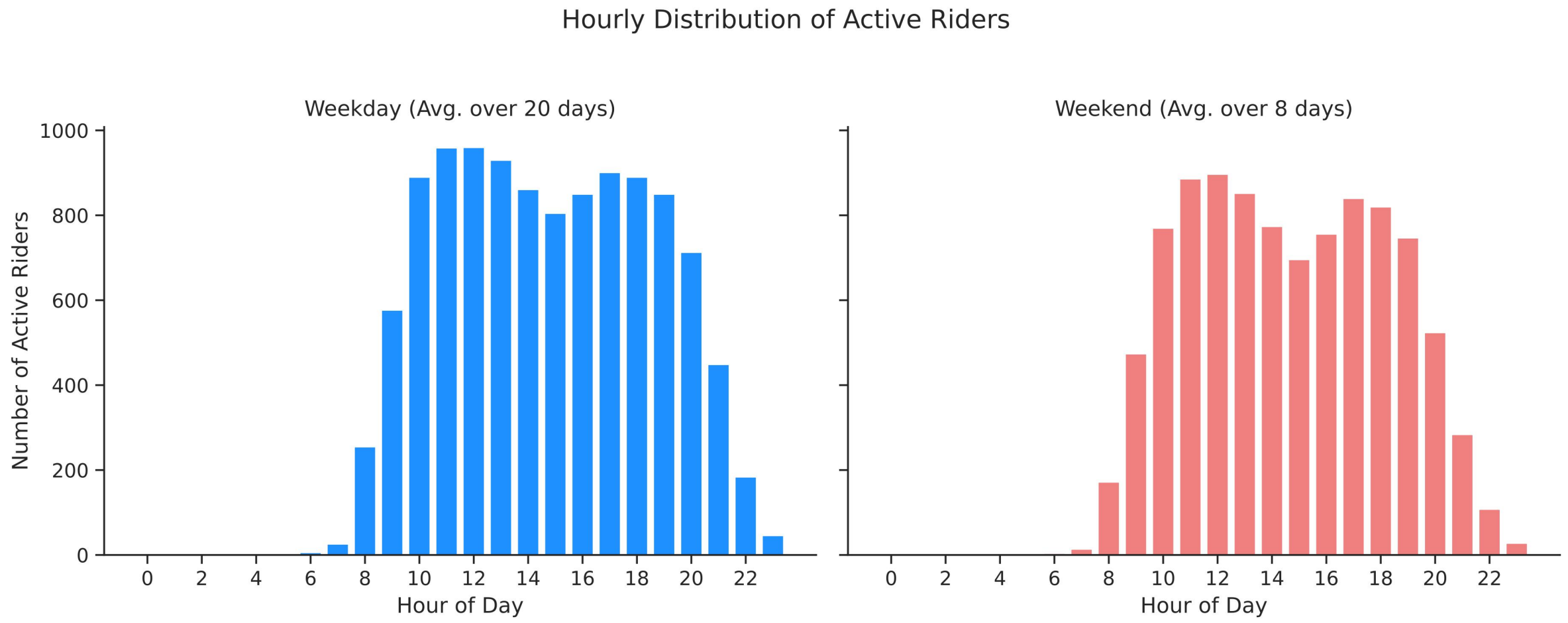

**Fig. S4 |** Average hourly distribution of active rider identifiers. The figure shows the average number of unique rider identifiers active in each hour, aggregated separately across 20 weekdays (blue, left) and 8 weekend days (red, right). Both panels indicate sustained rider activity through daytime and evening periods, with higher activity around lunch and dinner hours.

**Fig. S5** shows the hourly distribution of orders per active courier identifier. Across all reconstructed delivery waves, each wave contains an average of 3.36 orders. Weekday workload is more peaked around lunch and dinner periods, whereas weekend workload shows a more moderate fluctuation across daytime hours. These differences support the use of the dataset for studying source-record task rhythms and workload proxies, while direct inference about labor conditions requires complementary data.

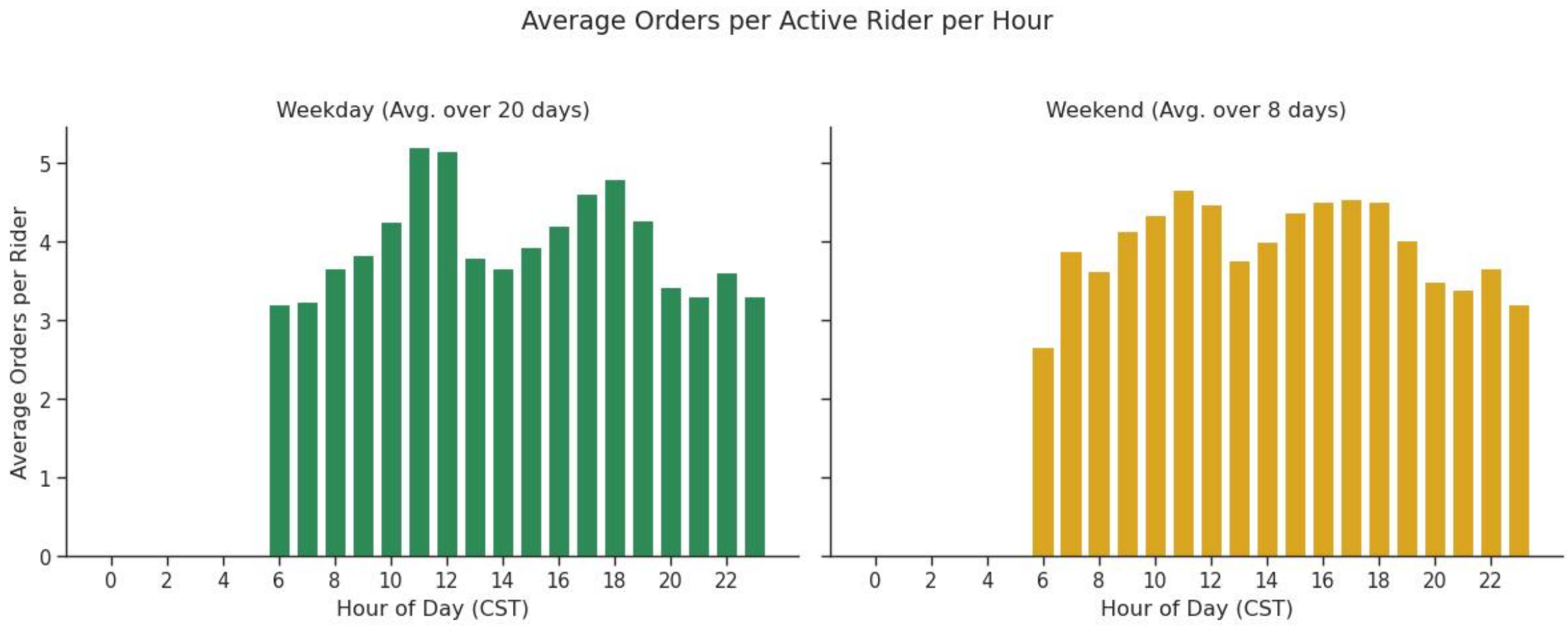

**Fig. S5** | Average orders per active rider per hour. This figure illustrates rider workload, measured as the average number of orders handled by an active rider in a wave per hour. The weekday panel (green, left) shows distinct peaks corresponding to lunch and dinner. The weekend panel (yellow, right) exhibits a more sustained plateau of high rider workload.